# The Role of Local Structure in the Enhanced Dynamics of Deformed Glasses

Entao Yang, Robert A. Riggleman

Department of Chemical and Biomolecular Engineering,
University of Pennsylvania, Philadelphia, PA 19104 USA

**Abstract:** External stress can accelerate molecular mobility of amorphous solids by several orders of magnitude. The changes in mobility are commonly interpreted through the Eyring model, which invokes an empirical activation volume whose origin remains poorly understood. Here, we analyze constant-stress molecular dynamics simulations and propose an extension of the Eyring model with a machine-learned field, softness. Our model connects the activation volume, an empirical parameter, to a structural property (softness). We show that stress has an inhomogeneous effect on the mobility that depends on local structure, which explains the narrower distribution of relaxation time observed under stress.

1. Introduction

Stress-induced dynamical acceleration at the molecular level, observed in both experiments [1–5] and simulations [6–8], is widely believed to be the origin of yielding and plastic motion in glasses [9–13]. In ductile polymers, this leads to tough materials that can dissipate large amounts energy before failure, enabling applications in protective coatings and membranes. As a result, understanding how external stress affects polymer dynamics is critical to understanding the mechanism of deformation.

The Eyring model is a classic generic model for describing the effect of external stress on dynamical yielding of activated processes, where an applied stress causes a linear reduction in the energy barrier impeding thermally activated motion [14]. This model and many later modifications [9,15,16] based on the idea that stress can induce faster mobility are still widely employed. However, many studies raise concerns about the validity of this model, mainly from two perspectives. (i) The effective free energy barrier is controlled by an empirical parameter, the activation volume, which has neither a clear physical meaning nor a connection to a particle-scale property. (ii) The thermal free energy barrier itself is assumed to be same everywhere, despite the difference in local structure and potential structural changes in preferred packing during deformation. Both of these concerns can be attributed to the key limitation of the Eyring model, where it does not have a clear structural dependence, neither on a microscopic level nor in an average sense.

It has been long suspected that structure plays an important role in glassy dynamics [17–19], though only recently have structural metrics that are truly predictive of glassy dynamics been developed [20]. Liu and coworkers recently developed a machine-learned field, softness, which characterize particles' local structure and shows a strong correlation with particle-level dynamics [21–23]. This new structural measurement has given insights to many aspects of glassy dynamics, including aging [24] to structural initiation of shear band [25,26]. With softness, bulk glassy dynamics can be decomposed as a product of two independent processes: one that depends on structure through softness, and one that is independent of softness [22,27]. These findings together delineate an inherent connection between structure and dynamics in glassy materials, which further emphasizes the potential for embedding structural components into the Eyring model.

In this study, we first expend the idea of dynamical decomposition in bulk glass [22,27] to systems undergoing creep deformation. We then prove its validity by proposing a structural-dependent Eyring model that applies within the linear and weakly nonlinear creep region. Our model connects the activation volume to a structural property (softness) quantitatively for the first time, accounts for the structural heterogeneity, and the model isolates the effect of stress from the structural evolution during deformation. We prove that the dynamical enhancement is heterogenous in the system; hard particles are enhanced more than soft particles, which explains the narrower distributions of relaxation time during deformation observed previously in both experiments [10,28,29] and simulations [11,29], even within the low strain regimes [2,30]. This is also consistent with the interpretation suggested by previous work, where dynamical acceleration induced by deformation is higher at slow regions [2].

## 2. Method

2.1 System details

We used a coarse-grained bead-spring model to construct the polymer matrix in our simulation system [31]. Each polymer chain consists of 128 Lennard-Jones (LJ) interaction sites, connected with fully flexible harmonic bonds. There are 405 chains in the system, yielding a total of 51,840 polymer particles. The standard 12-6 Lennard-Jones cut potential is used to describe all the non-bonded interactions,

$$U^{nb}(r_{ij}) = 4\varepsilon \left[ \left( \frac{\sigma}{r_{ij}} \right)^{12} - \left( \frac{\sigma}{r_{ij}} \right)^{6} \right] - U_{cut}, r_{ij} < 2.5\,\sigma$$

where $U_{cut}$ is the value of the 12-6 potential at our cut-off distance, $r_c = 2.5\,\sigma$. The polymer-polymer interaction ($\varepsilon_{pp}$) is fixed at 1.0. The bonded interactions are described by a harmonic bonding potential,

$$U_{ij}^{b} = K(r - \sigma)^2$$

where $K = 400\varepsilon/\sigma^2$, and $\sigma$ is the diameter of monomers. All the units are in LJ reduced units; the reduced temperature, $T$, is expressed as $T = kT^*/\varepsilon$, and the LJ time, $\tau_{LJ} = t^*\sqrt{\varepsilon/m\sigma^2}$, where $k$ is the Boltzmann constant, $m$ is the mass of a single LJ interaction site. $T^*$ and $t^*$ are temperature and time in laboratory units, respectively.

Systems are equilibrated in the NPT ensemble at $T = 1.0$ and $P = 0$ with a timestep of 0.002 $\tau_{LJ}^{-1}$. Connectively altering Monte Carlo moves are applied for the purpose of reaching equilibrium [32–34]. Independent configurations are separated by at least one polymer diffusion time, $\tau_D$, to guarantee their independence. Next, polymers are quenched to different target temperatures ($T = 0.35 - 0.42$) near our simulated $T_g = 0.46$ with a same cooling rate, $\Gamma = 10^{-4} \tau_{LJ}^{-1}$, followed by an aging process of 10,000 $\tau_{LJ}$. Creep deformations with a series of stresses ($\sigma_c = 0.1 \sim 0.5$) are then performed at different temperatures ($T = 0.35 \sim 0.42$). All the simulations are performed using the LAMMPS package [35].

2.2 Structural measurement: Softness

Softness is used to measure the polymer structure, which is a machine-learned field that shows strong correlation with particle-level dynamics [21–23]. Training softness begins by first identifying an equal number of 'soft' particles which are about to rearrange and 'hard' particles that go a long time without rearranging and characterizing their local environment using a group of $N$ (usually tens to hundreds) structure functions [21,36]. The values of these structure functions form a feature vector for every particle, and each particle corresponds to a point in a high dimensional space, $\mathbb{R}^N$. By applying a support vector machine (SVM) we can find a hyperplane that best separate the 'soft' particles and the 'hard' particles in $R^N$. The signed distance between the point and the hyperplane is defined as the 'softness' of the corresponding particle, which is positive for soft particles and negative for hard particles. Previous work has shown that the probability of observing a particle rearrange increases approximately exponentially with the particles' softness [21–23].

In this study, we use a hyperplane trained previously in a neat polymer system at T=0.50, where the only difference is polymer chain length. More technical details can be found in our recent work [26]. According to our tests, softness calculated with this hyperplane follows a normal distribution with a standard deviation $\sigma = 1.0$ and more than 90% of rearranging particles in our system have a positive softness (more details are provided in the ESI). For the following analysis, unless specified, we focus on the softness ranges from -3.5 to 1.5, covering 98% of particles.

2.3 Dynamical measurement: Particle's probability of rearrangement, $P_R$

Instead of the segmental relaxation time, $\tau_\alpha$, we use particles' probability of rearranging at a given softness, $P_R(S)$, as the measurement for monomer mobility. It has been shown that $\tau_\alpha$ can be predicted from knowledge of $P_R(S)$ and the softness distribution, where $\tau_\alpha \sim \frac{1}{P_R(<S>)}$ [22,24]. Materials below their glass transition undergo a process known as physical aging, whereby the material slowly evolves towards equilibrium with an increasing density and reduced mobility. It has been shown that deformation can alter aging dynamics and even reverse its effects (rejuvenation) [28,37,38]. Recent work has shown that the slow glassy dynamics during aging is a structural process (average softness decreases) and the $P_R - S$ relation remains unchanged for different aging times [24]. As a result, an essential advantage of our use of $P_R(S)$ to measure mobility is that we will be able to separate the effects of stress on the mobility from the structure changes that arise due to physical aging or rejuvenation.

3. **Results and discussion**

Recent works have shown that the glassy dynamics in the bulk glassy system [22] and polymer thin films [27] at a given temperature can be decomposed as a product of two independent processes as

$$P_R(S) = exp(\Sigma_0 - e_0/T) \times exp[(e_1/T - \Sigma_1)S] \equiv P_I(T)P_D(T,S). \qquad \text{Eqn. 1}$$

Here, $P_I(T)$ is structural-independent and $P_D(T,S)$ depends on structure through softness, and the four parameters $\Sigma_i$ and $e_i$ are independent of both temperature and softness. Here, we take the same protocol described in the literature [22,27] and start from the quiescent system at a series of T below $T_g$ ($T/T_g = 0.76 \sim 0.91$). As shown in Figure 1a and 1b, we observed a similar trend as bulk glass system [22], where particle's probability of rearrangement, $P_R$, exhibits an Arrhenius behavior at each softness value, and all the left-extended fitting curves share a common intersection point. The shared intersection point indicates that both $\Sigma$ and $e$ depend linearly on the softness and the polymer dynamics can be decomposed into two independent parts as shown above in Eqn. 1. The corresponding temperature of this intersection point has also been demonstrated to scale with the onset temperature of glassy dynamics [22].

We next test whether the dynamical decomposition procedure continues to apply for systems undergoing creep. We performed a series of constant uniaxial, tensile stress MD simulations with stress values ranging from 0.1 to 0.5 while maintaining constant pressure in the transverse directions. The strain-time curves and corresponding softness evolution at the highest temperature (T=0.42) can be found in Figure 1c and 1d (See results at other T in the ESI). Since our goal is to develop a structure based Eyring model, we primarily focus on the low stress regime where the Eyring model has been shown to be most effective [29]. More specifically, we use data collected from $t \in [0,2000]$ for $\sigma_c \in [0,0.40]$. While for $\sigma_c = 0.45$, we only use the time period of $t \in [0,1000]$. Both the strain and softness changes are larger at $\sigma_c = 0.50$ and larger deviation from the Arrhenius relation at each softness is observed when performing dynamical decomposition. Thus, we take $\sigma_c = 0.45$ as the upper limit for the decomposition in this work. The system average softness remains unchanged or increases steadily after the initial elastic response within this regime (See Figure 1f). Our results indicate dynamical decomposition still applies in polymer glasses under external stress, and we observe that the predicted onset temperature (given by the intersection point) shifts to lower values under deformation, which is consistent with the general notion that mobility can be enhanced by stress.

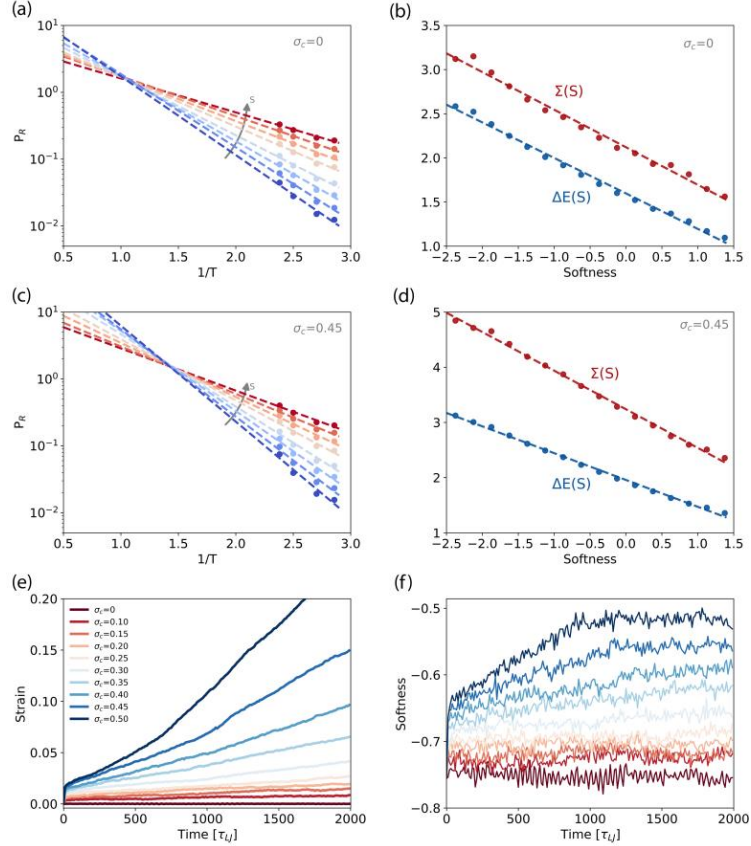

**Figure 1** Dynamical decomposition in glassy polymer: (a) Polymer monomer probability of rearrangement at a given softness, $P_R(S)$, as a function of $1/T$ at eight different softness values in the quiescent system ($\sigma_c = 0$). The color gradient represents the gradient in softness, ranging from dark blue at lowest S ($S = -2.75$) to dark red at highest S ($S = 1.25$). (b) $\Delta E$ (blue) and $\Sigma$ (red) as a function of softness, where $P_R = \exp(\Sigma(S) - \Delta E(S)/T)$. (c) $P_R(S)$ as a function of $1/T$ in polymers under creep where $\sigma_c = 0.45$. (d) $\Delta E$ and $\Sigma$ in polymers under creep where $\sigma_c = 0.45$. (e) Strain-time curves of polymers under creep at $T = 0.42$. (f) Evolution of average softness in polymers under creep at $T = 0.42$. The color gradient represents the gradient in stress, ranging from dark red for quiescent system ($\sigma_c = 0$) to dark blue at the highest stress ($\sigma_c = 0.50$).

The Eyring model is widely employed to describe the dynamical acceleration due to external stress in amorphous solids [9,10,14,29]. The segmental relaxation time at a certain applied stress, $\sigma_c$, can be predicted as $\tau_\alpha \propto \frac{\sigma_c}{\sinh[(\sigma_c \cdot V^*)/2k_BT]}$, where $V^*$ is an empirical activation volume and $k_BT$ is the thermal energy. Considering $P_R \propto \frac{1}{\tau_\alpha}$, we expect the changes in $P_R$ due to stress based on the Eyring model to be

$$P_R \propto \frac{\sinh\left[\frac{\sigma_c \cdot V^*}{2k_BT}\right]}{\sigma_c} = \frac{\exp\left[\frac{\sigma_c \cdot V^*}{2k_BT}\right] - \exp\left[-\frac{\sigma_c \cdot V^*}{2k_BT}\right]}{2\sigma_c}. \qquad Eqn.\ 2$$

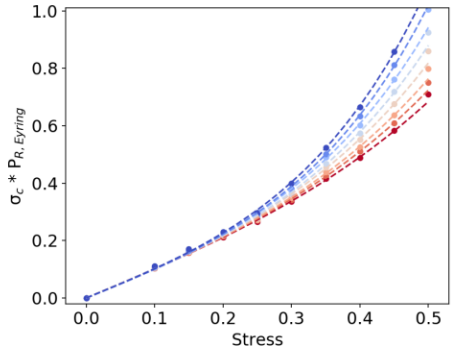

**Figure 2** $\sigma_c * P_{R,Eyring}$ as a function of stress at a given softness. Color gradient represents the same softness gradient presented in Figure. 1a and 1c. The dash curves are the fitting results for $\sigma_c \leq 0.45$.

To measure the dynamical enhancement induced by external stress, here we introduce a new term, $P_{R,Eyring} \equiv \frac{P_R(S)}{P_{R,u}(S)}$, which is the ratio of $P_R$ under creep over the undeformed system for a given softness. Inserting $P_{R,Eyring}$ into Eqn. 2, we have for the stress-induced acceleration

$$P_{R,Eyring} = A \times \frac{exp[\frac{\sigma_c \cdot V^*}{2k_BT}] - exp[-\frac{\sigma_c \cdot V^*}{2k_BT}]}{\sigma_c}. \qquad Eqn.\ 3$$

By rearranging terms and combining pre-factors, we obtain

$$\sigma_c * P_{R,Eyring} = A \times \left[ exp\left[\frac{\sigma_c \cdot V^*}{T}\right] - exp\left[-\frac{\sigma_c \cdot V^*}{T}\right]\right], \qquad Eqn.\ 4$$

where A is a constant pre-factor. In Figure 2, we plot $\sigma_c * P_{R,Eyring}$ as a function of stress for values of softness at T=0.42, and the measured dynamical enhancement can be nicely described by Eqn. 4 for stresses $\sigma_c \leq 0.45$. We choose this temperature because the statistics are usually better at high temperature since there are more rearrangements, especially for the hard particles, and other temperatures exhibit qualitatively similar behavior (see ESI for more details).

Results in Figure 2 suggest that the change of $P_{R,Eyring}$ agrees well with the Eyring model, which provides a route to look closely at the dynamical expressions of Eqn. 1 for glassy polymers under creep. To do that, we first need to address how do the parameters $A$ and $V^*$ depend on softness.

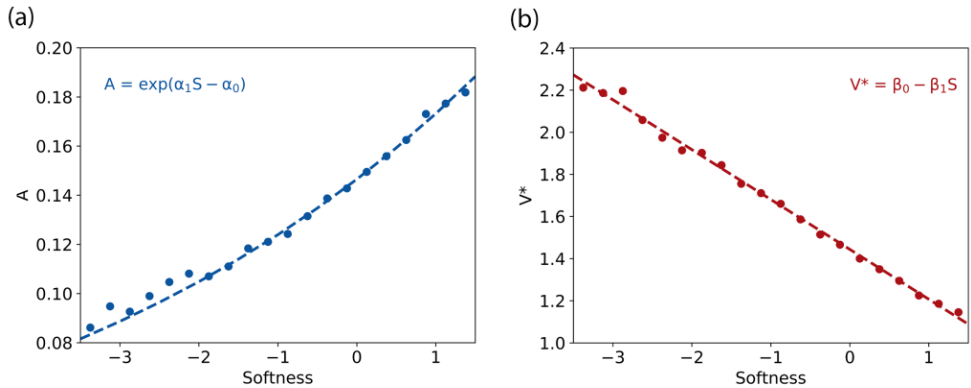

**Figure 3** Fitting parameters as a function of softness: (a) pre-factor $A$; (b) activation volume $V^*$.

In Figure 3, we plot $A$ and $V^*$ as a function of softness, finding that $A$ grows exponentially with $S$ while $V^*$ decreases linearly. As a result, we can write $A = \exp(\alpha_1 \cdot S - \alpha_0)$ and $V^* = \beta_0 - \beta_1 \cdot S$, where $\alpha_i$ and $\beta_i$ are independent of softness. We choose an exponential rather than a linear dependence for the pre-factor $A$ because (i) it is unphysical for $A$ to have a negative value and (ii) an exponential relation better matches the other terms in the dynamical decomposition model. Note that unlike the parameters ($\Sigma_i$ and $e_i$) we have in the original ($\sigma_c = 0$) model, we observe that both $\alpha_i$ and $\beta_i$ depend on T (see ESI for fitting at another temperature). This meets our expectation, because the magnitude of dynamical enhancement can vary strongly with T.

If we put the expressions for $A$ and $V^*$ into Eqn. 4 and combine with the above expression $P_R(S) = P_{r,u}(S) P_{R,Eyring}(S)$, we can arrive at a complete expression for polymer dynamics under creep,

$$P_R = \frac{\exp(\alpha_1 \cdot S - \alpha_0)}{\sigma_c} \cdot \left( \exp\left[\frac{\sigma_c \cdot (\beta_0 - \beta_1 \cdot S)}{T}\right] - \exp\left[-\frac{\sigma_c \cdot (\beta_0 - \beta_1 \cdot S)}{T}\right] \right) \cdot \exp(\Sigma_0 - \frac{e_0}{T}) \cdot \exp[-(\Sigma_1 - \frac{e_1}{T}) S], \quad \text{Eqn. 5}$$

which we refer to as the structure dependent Eyring model. Note that we have not taken any approximations beyond the functional forms of each term, so we expect this relationship to hold if we restrict ourselves within the low stress regime region, where both the Eyring model and decomposition of the dynamics into structure-dependent and independent terms remains valid.

Due to the existence of the hyperbolic sine term, it seems that we cannot as easily separate (decompose) the dynamics for systems under creep deformation into distinct structure dependent and independent terms. However, our simulations show that the negative exponential term decays rapidly with stress. In other words, when stress is not too small ($\sigma_c > 0.25$), Eqn. 5 can be written approximately as

$$P_R \cong \frac{1}{\sigma_c} \times \exp\left(\Sigma_0 - \alpha_0 + \frac{\beta_0 \cdot \sigma_c - e_0}{T}\right) \times \exp\left[-\left(\Sigma_1 - \alpha_1 + \frac{\beta_1 \cdot \sigma_c - e_1}{T}\right) S\right]. \quad \text{Eqn. 6}$$

From Eqn.6 we can see that $P_R$ can be approximated as a product of two exponential terms when external stress is large enough, where one depends on softness and the other does not. This makes $P_R$ have a similar form as Eqn.1, explaining why dynamical decomposition works in this situation.

In the limit of small applied stress, $\frac{\sigma_c \cdot (\beta_0 - \beta_1 \cdot S)}{T}$ is close to 0, and a Taylor expansion for the sinh term yields $2 * \frac{\sigma \cdot (\beta_0 - \beta_1 \cdot S)}{T}$, which is not an exponential term and appears to preclude dynamical decomposition. However, when stress is small the dynamical change is also insignificant and $P_{R,\ Eyring}$ is nearly 1, allowing us to assume the effects of stress can be decomposed. Thus, the dynamical decomposition is valid in both cases, consistent with the results presented in the last section.

It is long believed that the activation volume in the Eyring Model is an empirical fitting parameter and not correlated with any polymer structural properties [9,10,16]. Traditionally, the activation volume $V^*$ is a constant for a given material at certain conditions (like temperature). However, our results indicate both that $V^*$ is not constant but heterogenous in glassy polymers, and furthermore it decreases linearly with softness, a measurement of particle's local structure. For the harder particles, which are more difficult to rearrange even under deformation, the activation volume is larger, corresponding to a larger energy barrier decrease for the hopping. Integrating $V^* = \beta_0 - \beta_1 \cdot S$ over the softness distribution at different stress yields an overall $V^*$ similar to traditional values, $V^* \approx 1.6$, which agrees with the interpretation that $V^*$ is the cooperative movement of two to three segments [10]. Our results also show that this overall $V^*$ depends weakly on stress, since external stress can slightly increase the mean softness.

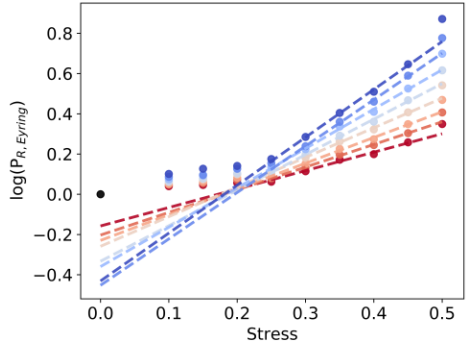

**Figure 4** $\log P_{R,Eyring}$ as a function of stress at a given softness. Color gradient represents the same softness gradient presented in Figure 1a and 1c (Fit for $0.3 \leq \sigma_c \leq 0.45$, excluded the last point).

Based on our derivation above, there are two different regimes for polymer dynamical enhancement under creep. When the stress is sufficiently small (i.e., the two stress-involved exponential terms are in the same order of magnitude), the dynamical acceleration is relatively small, leading to a weak dependence on particles' local environment. When the stress becomes larger (but not so large as to induce yielding and an increasing strain rate), polymer dynamics can be expressed as a product of two independent processes for a given stress. In Figure 4, we plot the $\ln P_{R,Eyring}$ as a function of stress at T=0.42, and we find a turning point that indicates this crossover in behavior at approximately $\sigma_c = 0.25$. Below this stress, the dynamical enhancement is relatively small, although there is still a weak dependence on the structure. After this point, the enhancement grows exponentially with stress, and the growth is larger for the hard particles. This threshold value and unsmoothed dependences of the mobility on stress has been observed in experiments [1].

The fitting curves of different softness share a common intersection point, which is expected from the linear relation between activation volume and softness. In other words, for polymer glass under creep the dynamical enhancement directly depends on the local structure, and the mobility acceleration is larger for the hard particles compared with soft particles. While this heterogenous acceleration offsets the heterogenous structure in the polymer glass, the structure heterogeneity remains unchanged during creep, as evidenced by the constant variation of softness distribution (see ESI for more details). This ultimately leads to a narrower distribution of mobility for polymer under external stress. We observe small deviations from the model at $\sigma_c = 0.5$, which approximately corresponds to where the strain rate starts increasing.

## 4. Conclusion

In summary, we have demonstrated that in glassy polymer systems the dynamics can be decomposed into two independent processes, and this decomposition can be further expanded to systems under creep, at least within the low stress regime where Eyring model is valid. By introducing a dynamical enhancement ratio, $P_{R,Eyring}$, we show that the activation volume, $V^*$, decreases linearly with softness, which is a measurement of particles' local structure. To the best of our knowledge, this is the first time that the activation has been correlated with a structural property. Our model also reveals that there is a turning point for the dynamical enhancement for polymer glass under external load, as predicted by the structural-dependent Eyring model. When the stress is small, the acceleration is small and the dependence on structure is weak. When it is sufficiently large enough and make the negative exponential term in Eyring model negligible, the dynamical enhancement grows exponentially with stress and shows a strong dependence on the softness. The mobility of hard, better-packed particles accelerates more compared to the soft particles. This heterogenous dynamical enhancement reduces the dynamical heterogeneity caused by the structure, which remains unchanged under stress, and leads to the narrower mobility distribution observed in polymer glass under active deformation.

## Acknowledgements

We thank Karen Winey, Bharath Natarajan, and James Pressly for helpful discussion. This work was supported by ExxonMobil Research and Engineering as well as DOE-BES via DE-SC0016421. We also acknowledge computational resources provided through XSEDE allocation TG-DMR150034.

# The Role of Local Structure in the Enhanced Dynamics of Deformed Glasses

## Supplementary Materials

Entao Yang and Robert Riggleman*

Department of Chemical and Biomolecular Engineering,

University of Pennsylvania, Philadelphia, PA 19104

1. **Softness distribution of the pre-deformation system at T=0.50 ($T/T_g = 1.09$).**

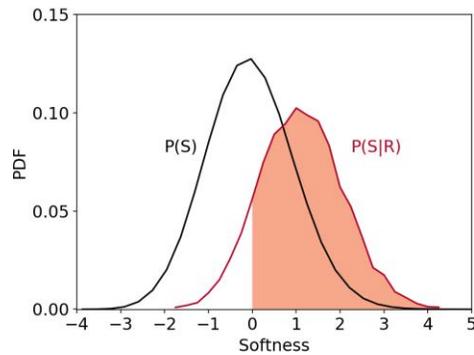

**Fig. S1** Softness distribution of the pre-deformation system at T=0.50. $P(S)$ is the distribution of all the particles (black) and $P(S|R)$ is the distribution of the rearranging particles (red). More than 90% of the rearranging particles have a positive softness (represented by the shaded region), signifying the validity of our hyperplane.

## 2. Strain-time curves and softness-time curves at T=0.35 ($T/T_g = 0.76$).

In Fig. S2, we plot the strain and the average softness as functions of time for systems under different stresses (0~0.50) at T=0.35 ($T/T_g = 0.76$), which is the lowest temperature we have for this study. These curves show that within the time period we defined in the main text (linear response region), both strain and softness grow steadily during deformation, after the initial elastic response. This trend is same as the one shown in **Fig. 1** in the main text and guarantee the prerequisite (linear response region) we made for the dynamical decomposition.

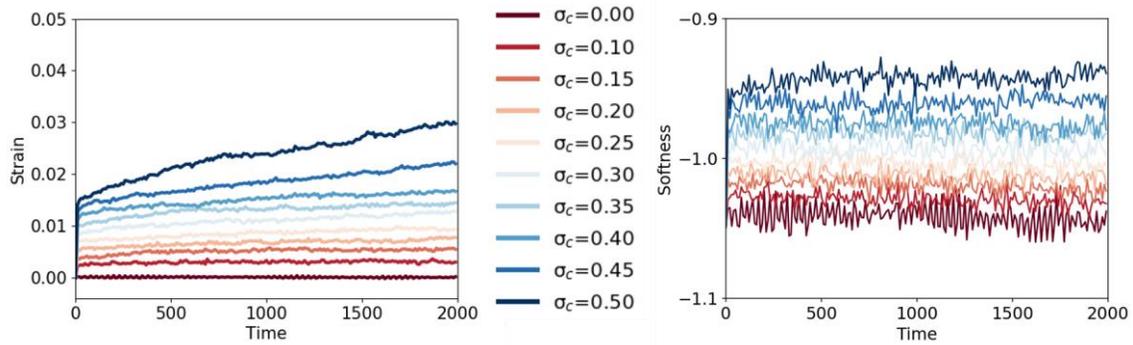

**Fig. S2 Left,** strain-time curves at T=0.35; **Right**, softness-time curves at T=0.35. The color gradient represents the gradient in stress, ranging from dark red for quiescent system ($\sigma_c = 0$) to dark blue at the highest stress ($\sigma_c = 0.5$).

## 3. Dynamical enhancement predicted with the Structural-dependent Eyring Model at T=0.40 ($T/T_g = 0.87$)

In **Fig. S3**, we performed the same analysis for the Structural-dependent Eyring Model at T=0.40. We choose this temperature because enhancement at lower T are smaller and statistics becomes poor since there are fewer rearrangements for the low softness particles. These results further verify that the structural-dependent Eyring model can predict the dynamical enhancement for particles with different softness. The activation volume $V^*$ decreases linearly with S, while the pre-factor A holds an exponential relation with S.

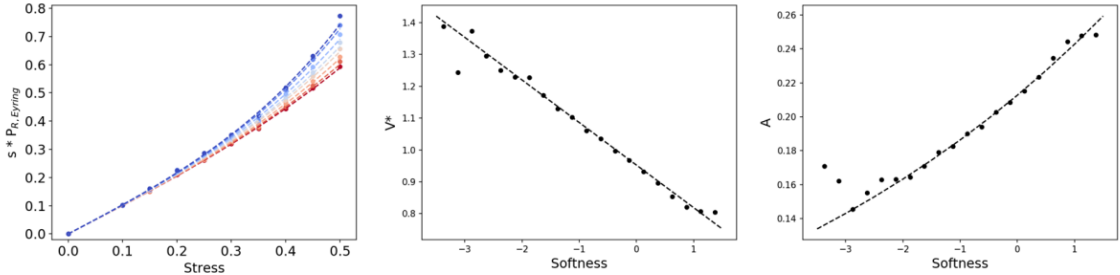

**Fig. S3** Prediction with the Structural-dependent Eyring Model at T=0.40. **Left,** $\sigma_c * P_{R, Eyring}$ as a function of stress at a given softness. Color gradient represents the same softness gradient presented in **Fig. 1** in the main text. The dash curves are the fitting results for $\sigma_c \leq 0.45$. **Center**, activation volume $V^*$ as a function of softness. **Right**, pre-factor A as a function of softness. Dash curves are the fitting results for $S \geq -2.0$, because there are few rearrangements for smaller S at this temperature.

4. **Softness distribution under external stress**

   In **Fig. S4,** we plot the softness distribution under different external stresses ($\sigma_c \leq 0.50$) at T=0.42, which is the highest temperature we have in this study. Results indicate that softness distribution remains almost unchanged, where only the mean softness increases with stress. This suggests that the structural heterogeneity remains unchanged during creep deformation.

   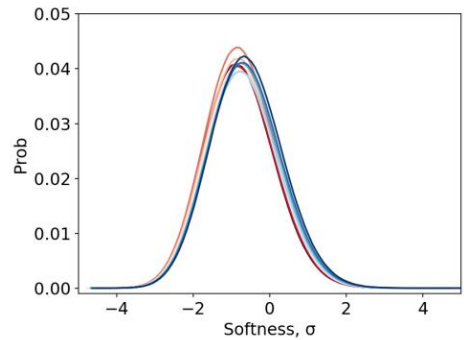

   **Fig. S4** Softness distribution in glassy polymer under different external stresses. The color gradient represents the same gradient in stress shown in **Fig. S2**.